\documentclass[aps,prl,floats,floatfix,superscriptaddress,twocolumn]{revtex4-1}
\usepackage{ifthen}
\usepackage{ifpdf}
\usepackage{color}
\ifpdf
\usepackage{graphicx}
\usepackage{epstopdf}
\else
\usepackage{graphicx}
\usepackage{epsfig}
\fi
\usepackage{ulem}
\usepackage{latexsym}
\usepackage{amsmath}
\usepackage{amssymb}
\usepackage{bm}
\usepackage{wasysym}

\usepackage{latexsym}
\usepackage{amsmath}
\usepackage{amssymb}
\usepackage{bm}
\usepackage{wasysym}
\usepackage{ulem}
\usepackage{braket}
\usepackage{hyperref}

\begin{document}
	
	\title{Localization and delocalization in networks with varied connectivity}

\author{Tamoghna Ray}
\email{tamoghna.ray@icts.res.in} 
\address{International Centre for Theoretical Sciences, Tata Institute of Fundamental Research, Bengaluru -- 560089, India}

\author{Amit Dey}
\email{amit.dey.85@gmail.com }
\address{Ramananda College, Bankura University, Bankura 722122, India}
\address{International Centre for Theoretical Sciences, Tata Institute of Fundamental Research, Bengaluru -- 560089, India}

\author{Manas Kulkarni}
\email{manas.kulkarni@icts.res.in}
\address{International Centre for Theoretical Sciences, Tata Institute of Fundamental Research, Bengaluru -- 560089, India}

	\date{\today}

\begin{abstract}
    We study the phenomenon of localization and delocalization in a circuit-QED network with connectivity varying from finite-range to all-to-all coupling. We find a fascinating interplay between interactions and connectivity. In particular, we consider (i) Harmonic (ii) Jaynes-Cummings and (iii) Bose-Hubbard networks. We start with the initial condition where one of the nodes in the network is populated and then let it evolve in time. The time dynamics and steady state characterize the features of localization (self-trapping) in these large-scale networks. For the case of  Harmonic networks, exact analytical results are obtained and we demonstrate that all-to-all connection shows self-trapping whereas the finite-ranged connectivity shows delocalization. The interacting cases (Jaynes-Cummings, Bose-Hubbard networks) are investigated both via exact quantum dynamics and semi-classical approach. We obtain an interesting phase diagram when one varies the range of connectivity and the strength of the interaction. We investigate the consequence of imperfections in the cavity/qubit and the role of inevitable disorder. Our results are relevant especially given recent experimental progress in engineering systems with long-range connectivity. 
\end{abstract}

\maketitle


\textit{Introduction: }
The phenomenon of macroscopic self-trapping has been a subject of great interest and has shown to occur in a variety of systems, both theoretically and experimentally. Notable examples include bosonic Josephson junctions (BJJs) consisting of cold-atomic Bose-Einstein condensates (BECs) \cite{milburn1997quantum,raghavan1999coherent, smerzi1997quantum,levy2007ac, albiez2005direct,xhani2020dynamical, o2012quantum} and photonic systems \cite{shelykh2008josephson, makin2009time, abbarchi2013macroscopic, coto2015self, schmidt2010nonequilibrium, raftery2014observation} characterized by light-matter interactions. Dissipative effects were also considered and delocalization - localization transition of photons was theoretically predicted \cite{schmidt2010nonequilibrium} in dissipative quantum systems and subsequently experimentally observed \cite{raftery2014observation}.

Although self-trapping was achieved, the inevitable photon leakage and spontaneous decay of the qubit limit the longevity of self-trapped states in realistic systems. To circumvent this, a drive was recently introduced and it was shown theoretically that a delicate interplay between drive, dissipation, interaction and kinetic hopping can lead to indefinitely long lived self-trapped states \cite{dey2020engineering}. This was an important step forward as it provides a protocol to ensure indefinitely long lived localization inspite of cavity/qubit imperfections.

\begin{figure}[h]
    \centering
    \includegraphics[scale = 0.23]{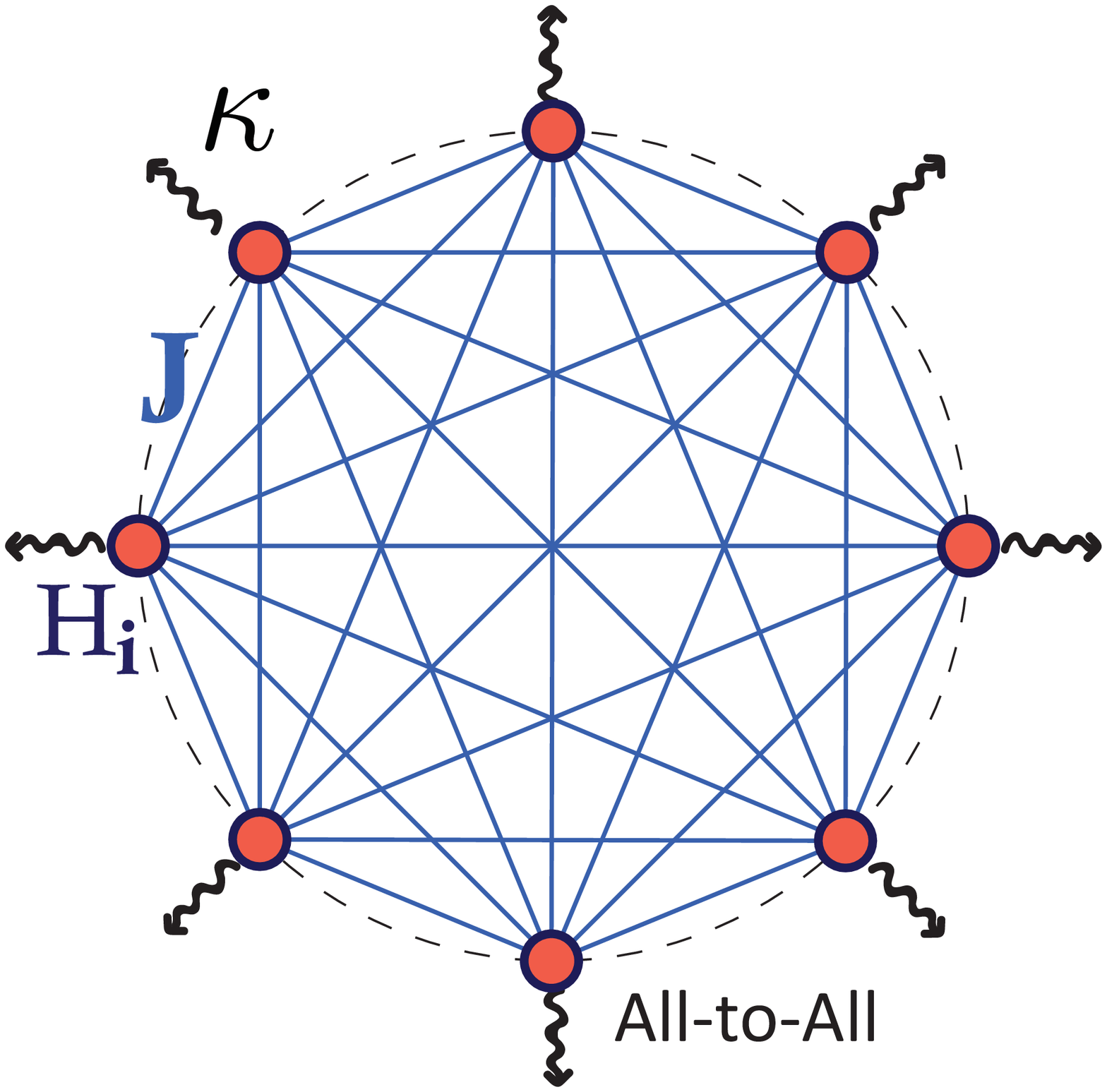}
    \includegraphics[scale = 0.23]{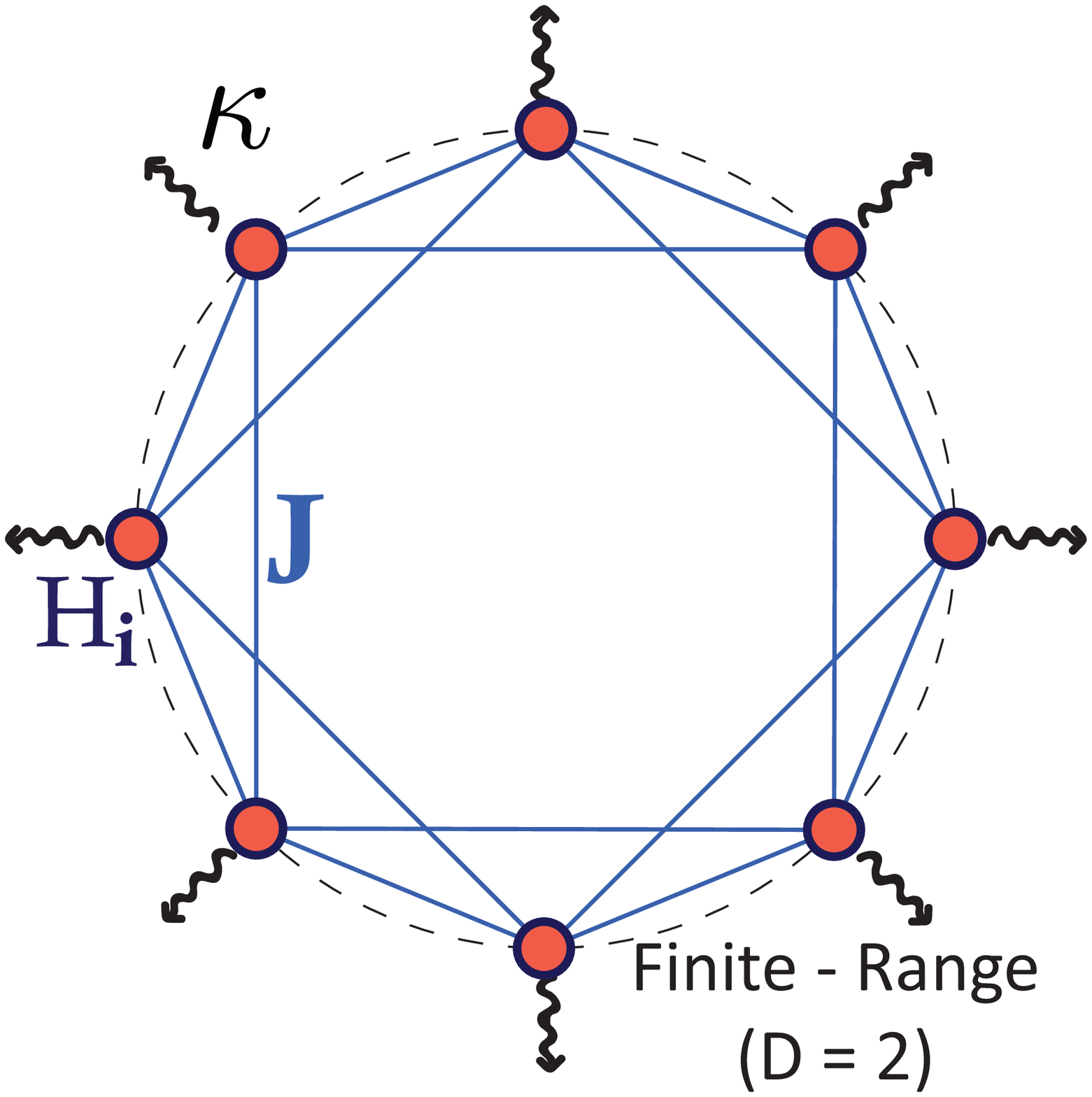}
    \caption{A schematic representation of the circuit-QED network considered in this work. The solid red circles denote the $N$ units ($H_i$ where we label $i=0,1\dots N-1$), the blue lines denote the inter-unit coupling with coupling strength $J$. Each unit is connected to its respective reservoirs (which can be characterized by cavity dissipation rate $\kappa$, qubit decay $\gamma$ and qubit dephasing $\gamma_\phi$). The figure on the left indicates a network with all-to-all coupling and the figure on the right indicates finite-range coupling with $D=2$ nearest neighbours. One can note from the figure on the right that if $D\geq \lceil N/2 \rceil$ (where $\lceil  \,\rceil $ denotes the ceiling function) then the unit effectively couples to all other units which makes it fall under the all-to-all case.}
    \label{schematic}
\end{figure}

While most of the works above were either restricted to dimer systems or 1D arrays, there is an important gap that needs to be addressed for the case when one has a highly non-trivial network of interacting systems. The properties of the network (such as connectivity) is expected to have an interesting impact on the phenomenon of localization and delocalization. This line of investigation is particularly important given recent experimental advances \cite{wright2019benchmarking, song2019generation, xu2020probing, hazra2021ring} in designing networks with varied connectivity (from finite-range to long-range) and success in establishing coupling between distant qubits \cite{xu2019cavity, borjans2020resonant, ritter2012elementary}. The recent experimental designs are also amenable to enhanced magnitude of connectivity. Additionally, such designs are tunable \cite{vaidya2018tunable}, scalable \cite{giovannetti2000scalable,fitzpatrick2017observation,arute2019quantum,ritter2012elementary}, generalizable to a wider range of platforms and potentially relevant for quantum computation \cite{kimble2008quantum,dey2015decoherence,li2018one}, modeling artificial light harvesters \cite{chin2010noise,cheng2009dynamics}, and quantum simulation schemes.

In this work, we consider a circuit-QED network with varied connectivity which is shown in the schematic Fig.~\ref{schematic}.
We consider a general set-up where each unit is connected to other units (nodes) via a  hopping term $J$ that is preferably uniform between units. The connectivity may extend to a finite number of neighbours [Fig.~\ref{schematic} (right)] or to all of them [Fig.~\ref{schematic} (left)].

Each unit could in principle be any local Hamiltonian ($H_i$) and we consider (i) Harmonic (ii) Jaynes-Cummings (JC) and (iii) Bose-Hubbard (BH) model. Our main results can be summarized as follows - (i) For Harmonic networks, we present exact analytical results and highlight the clear distinction between finite-range and all-to-all coupling scenarios in terms of degree of photon localization. Here, one would naively expect that an excited unit of a harmonic network is prone to lose its excitation as its connectivity with the rest of the units increases. Surprisingly, this expected feature is contradicted by our observation. (ii) For the JC and BH networks, by both exact quantum and semi-classical approach, we show the intricate interplay between connectivity and interactions, demonstrating self-trapped and delocalized regimes. (iii) Interesting non-monotonic features in the degree of localization are observed as one changes the number of units. (iv) The role of cavity dissipation, qubit decay/dephasing and inevitable disorder has been highlighted. 

\textit{The Hamiltonian: }
We consider a system of $N$ units $H_i$ as shown in  Fig.~\ref{schematic} which are connected to one another with a hopping term $J$. The Hamiltonian for such a system when there is all-to-all coupling is given by
\begin{equation}
    H = \sum_{i=0}^{N-1} H_i -\frac{J}{2} \sum_{\substack{i,j = 0 \\ i\neq j}}^{N-1} \left(a^\dagger_i a_j + h.c.\right),
    \label{Hamil}
\end{equation}
where $H_i$ is the Hamiltonian for the $i^{\rm th}$ unit. Although, in Eq.~\eqref{Hamil} we have considered all-to-all coupling, later on we will also present results for finite neighbour coupling. In this study, we explore three different cases - Harmonic, JC and the BH network. In all these cases, bearing physical systems in mind, our objective is to study networks with varied sizes and connectivity keeping the hopping strength and total number of excitations fixed.

\begin{figure}[htb]
    \centering
    \includegraphics[scale = 0.38]{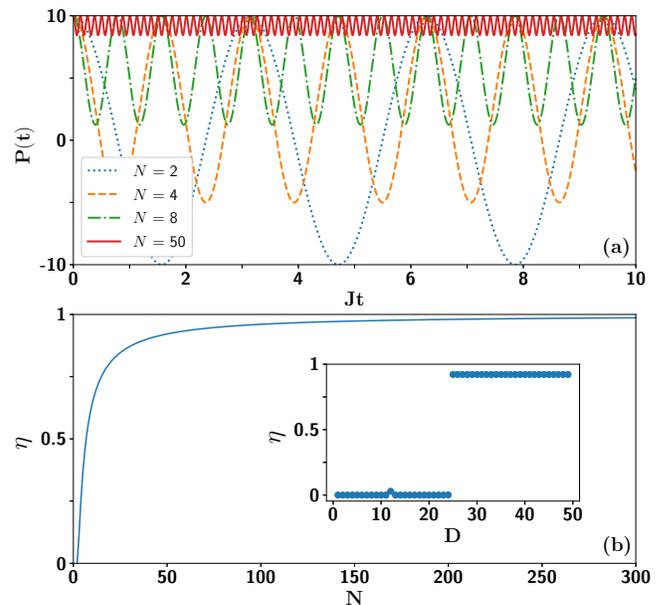}
    \caption{(a) $P(t)$ obtained in Eq.~(\ref{P_AtoA}) as a function of time (in units of $1/J$) for a Harmonic network for varying number of units with all-to-all coupling. We prepare the system initially such that there are $N_{\rm p} = 10$  bosons in the test unit. 
(b) $\eta$ obtained in Eq.~(\ref{eq:eta_aa}) as a function of $N$ is presented, clearly indicating stronger localization for large system sizes. The inset in Fig.~\ref{section1}(b) shows $\eta$ as a function of the number of nearest neighbours $D$ for $N = 50$, illustrating the loss of localization on losing the all-to-all configuration, {i.e., when $D <  \lceil N/2 \rceil$} .}
    \label{section1}
\end{figure}

\textit{Harmonic Network:}
In this case,
   $ H_i = \omega_c a_i^\dagger a_i $,
where $\omega_c$ is the cavity frequency and $a_i^\dagger \text{ and }a_i$ are the creation and annihilation operators for the photons in the $i^{\rm th}$ cavity. Such a system is exactly solvable, and we obtain analytical results for this network. We prepare the system in an initial state such that there are $N_{\rm p}$ bosons in the test unit ($i=0$) and all other units are empty. The system is let to evolve in time and one can compute observables such as the bosonic occupation at the $i^{\rm th}$ unit, $n_i (t) = \langle a_i^\dagger (t) a_i  (t) \rangle$ where $\langle \rangle$ is the quantum mechanical average. To study the dynamics of the bosons we define the quantity $P (t) = n_0 (t) - \sum_{j = 1}^{N-1}n_j  (t)$, which gives the population difference between the test unit and the rest of the system (imbalance). We introduce a convenient diagnostic tool to compute the degree of localization,
\begin{equation}
    \eta = \frac{N_{\rm p} + P_{\rm min}}{2N_{\rm p}},
    \label{eq:eta}
\end{equation}
where $P_{\rm min}$ is the minimum value that $P(t)$ takes during its entire time evolution. Complete localization is characterized by $\eta \to 1$ and complete delocalization (i.e., all the photons have left the test unit) is characterized by $\eta \to 0$. For other values ($0 < \eta < 1 $), the quantity $\eta$ indicates the degree to which bosons in the test unit stay trapped. 

To obtain the evolution of $P(t)$ we calculate the equations of motion (EOM) of the system operators. For all-to-all coupling, we have (setting $\hbar =1$)
\begin{equation}
    \dot{a}_i = i [H, a_i] =  -i \omega_c a_i + i J \sum_{\substack{j  = 0 \\(j \neq i)}}^{N-1} a_j,
    \label{eom_AtoA}
\end{equation}
and for $D$ finite neighbour coupling (note that $D< \lceil N/2 \rceil$), we have
\begin{equation}
    \dot{a}_i = -i \omega_c a_i + i J \sum_{d = 1}^D\left(a_{i+d} + a_{i-d}\right).
    \label{eom_finite}
\end{equation}
When $D\geq \lceil N/2 \rceil$ we have the same setup as for all-to-all coupling. Eq.~\eqref{eom_AtoA} and Eq.~\eqref{eom_finite} can be written as $\dot{\textbf{x}} = \textbf{Ax}$, where $\textbf{x} = \left[a_0 \ a_1 \dots a_{N-1}\right]^T$. This can be solved by evaluating the eigenvalues ($\{\lambda_\alpha\}$) and eigenvectors ($\{ \textbf{u}_\alpha \}$) of $\textbf{A}$ ($\alpha = 1,2... N$), i.e., 
$    \textbf{x(t)} = \sum_{\alpha=1}^N c_\alpha e^{\lambda_\alpha t}\textbf{u}_\alpha $ where $c_\alpha$ are the weights of the initial condition on the eigenvectors. 
We formulate an analytical solution for $P(t)$ by shifting to the Fourier space, $a^\dagger_j = \frac{1}{\sqrt{N}}\sum_{k=0}^{N-1} e^{-i \frac{2\pi}{N} kj}a^\dagger_k$. In this space, the Hamiltonian is diagonal and $P(t)$ for $D$ finite neighbours becomes \cite{SM}

\begin{equation}
    \begin{gathered}
        P(t) = \frac{2N_{\rm p}}{N^2}\Bigg\{ 1 + \sum_{k,k'=1}^{N-1}\exp\left[i2Jt\left( f(k') - f(k) \right)\right]\\ + 2\sum_{k=1}^{N-1}\cos\left[  2Jt\left(f(k) -D \right) \right] \Bigg\}- N_{\rm p}, 
        \label{eq:PfiniteD}
    \end{gathered}
\end{equation}
where
\begin{equation}
    f(k) = \frac{\cos\left( \frac{D+1}{N}\pi k\right)\sin\left(\frac{D\pi k}{N} \right)}{\sin \left( \frac{\pi k}{N} \right)}.
\end{equation}
For all-to-all coupling, we get
\begin{equation}
    P(t) = \frac{N_{\rm p}}{N^2}[1 + (N-1)(N + 4\cos (NJt) -3)].
    \label{P_AtoA}
\end{equation}
From Eq.~\ref{P_AtoA} and Eq.~\ref{eq:eta},  we get 
\begin{equation}
\eta = 1 -\frac{4}{N}+\frac{4}{N^2} \quad \text{(all-to-all coupling),} 
\label{eq:eta_aa}
\end{equation}
which implies that we achieve stronger localization of photons in the test unit as we increase the system size $N$. 
   Note that $\lim_{N\to \infty} \eta = 1$ (for all-to-all coupling case). 
In Fig.~\ref{section1}(a) we plot $P(t)$ as a function of time (in units of $1/J$). We notice that when the number of units are small there is delocalization (accompanied by oscillations). However, upon increasing the number of units the network tends to get more localized and eventually there is perfect self-trapping in the large-$N$ limit. Fig.~\ref{section1}(b) shows the degree of localization ($\eta$) as one increases in number of units which is given by Eq.~\ref{eq:eta_aa}. {The inset in Fig.~\ref{section1}(b) demonstrates the loss in self-trapping in the finite-range case.} 

It is worth noting that,  alternatively we can numerically compute the entire correlation $N\times N$ matrix $C(t) = \langle \textbf{x} (t) \textbf{x}^{\dagger}(t) \rangle$ which is given by $C(t) = e^{-iht} C(0)  e^{iht}$ where $h$ is the single particle Hamiltonian ($N \times N$ matrix which contains information of whether the network connectivity is finite-range or all-to-all) that appears in $H = \sum_{i,j=0}^{N-1} h_{ij} a^\dagger_i a_j $. From $C(t)$, we can extract $P(t) = 2C_{00}(t)-\rm{tr}[C(t)]+(N-2)$ and this is in perfect agreement with our analytical expressions derived above [Eq.~(\ref{eq:PfiniteD}) and Eq.~(\ref{P_AtoA})]. Next, we discuss the case of {onsite} interactions/anharmonicity.

\textit{The Jaynes-Cummings Network: }
One can envision a situation where each cavity hosts a qubit. This leads to the well known Jaynes-Cummings system whose Hamiltonian is given by
\begin{equation}
    H_i = \omega_c a_i^\dagger a_i + \omega_q \sigma_i^+\sigma_i^- + g (a_i \sigma_i^+ + a_i^\dagger \sigma_i^-),
  \label{JCM}
\end{equation}
where $\omega_c$ and $\omega_q$ are the cavity frequency and the energy gap of the qubit respectively, g is the cavity-qubit coupling strength and $\sigma_i^+$ and $\sigma_i^-$ are the raising and lowering operators for the $i^{\rm th}$ qubit respectively. For the case of $N=2$ (dimer), delocalization - localization transition was investigated theoretically \cite{schmidt2010nonequilibrium} and was subsequently observed {experimentally}  \cite{raftery2014observation}. Recently, the intricate role of drive to circumvent inevitable imperfections was investigated for the JC dimer and 1D array leading to the proposal of indefinitely long lived self-trapped state \cite{dey2020engineering}. Given recent experimental advances in designing {highly connected} networks (which is a non-trivial generalization of dimer), we consider the setup with long-range connectivity. Each unit $H_i$ is described by Eq.~\eqref{JCM} and these units sit in networks such as the one shown in Fig.~\ref{schematic}. The role of light-matter interaction $g$ and the number of neighbours $D$ on the dynamics in such networks is investigated {for the resonant case ($\omega_c = \omega_q $)}. The system therefore is an ideal platform for understanding the intricate interplay between network connectivity ($D$), interactions ($g$) and kinetic hopping ($J$).

\begin{figure}[htb]
    \centering
    \includegraphics[scale = 0.40]{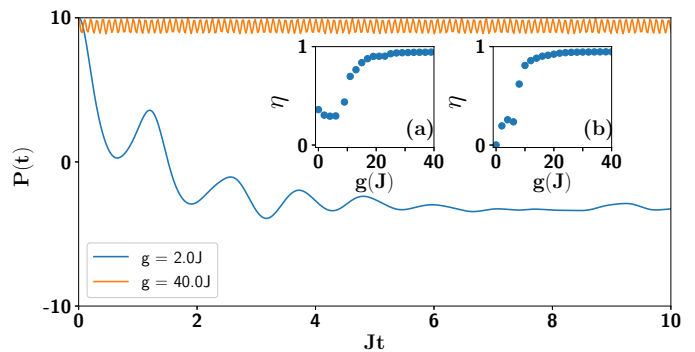}
    \caption{{The population imbalance $P(t) = n_0 (t) - \sum_{j = 1}^{N-1}n_j  (t)$} as a function of $t$ (in units of $1/J$) is plotted for a network of $N=5$ JC units (resonant case $\omega_c = \omega_q$) for different values of $g$ using exact quantum numerics. This is the case of all-to-all coupling where the initial condition is such that there are $N_{\rm p} =10$ photons in the test unit and rest of the units are empty. Qubits in all units are initially assumed to be in the ground state. It is evident from the figures that for weak interactions ($g$) one gets delocalization and we reach a self-trapped state when we increase the interaction. Inset (a) and (b) show the degree of localization as a function of $g$ for all-to-all coupling and nearest-neighbour coupling cases respectively.}
    \label{section2a}
\end{figure}

\begin{figure}[htb]
    \centering
    \includegraphics[scale = 0.6]{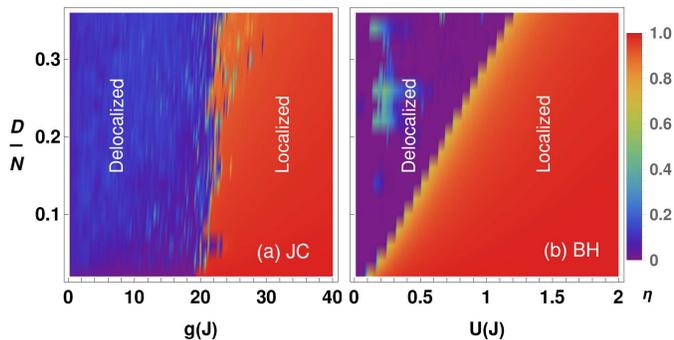}
    \caption{Heat map for $\eta$ with varying the number of nearest neighbours $D$ (as a fraction of the total number of units) and interactions for (a) JC and (b) BH using the semi-classical approach. The parameters are $N = 50$ and $N_{\rm p} = 50$. 
    Results are plotted for upto $18$ nearest neighbours. For the JC network the resonant case, i.e., $\omega_c = \omega_q$, is considered. We find a transition from delocalized to self-trapped regime.}
    \label{fig:heatmap}
\end{figure}

We present an exact quantum solution for Eq.~\eqref{eom_AtoA} where $H$ now stands for the JC network. The system is initially prepared with all the excitations ($N_{\rm p}$) in the cavity mode of the test unit and all other cavities and qubits are kept in the ground state, i.e., $ n_0(0) = N_{\rm p}$, $n_{i\neq 0}(0)= 0$ and $\langle \sigma^z_i(0) \rangle = -1$. Since there is no drive in the system, the oscillators can never be excited beyond $\ket{N_{\rm p}}$ (in each unit) and hence we truncate the basis at $\ket{N_{\rm p}}$. The dimension $\mathcal{N}_d$ of this truncated Hilbert space for the entire system is given by,
    $\mathcal{N}_d = \left[2 (N_{p}+1)\right]^{N}$.
For $5$ units and $10$ photons, $\mathcal{N}_d = 22^5$ (larger than $2^{22}$) and we numerically evolve the system with the Hamiltonian for these parameters. In  Fig.~\ref{section2a}, we investigate $P(t)$ as a function of time. One can notice that for small $g$  the photons delocalize and for large $g$, the photons are localized in the test unit. This self-trapped phenomenon for large $g$ is a result of an interesting interplay between interaction, kinetic hopping and network connectivity. Fig.~\ref{section2a} has two insets. The inset (a) shows the degree of localization as one tunes the interaction strength $g$ for all-to-all coupling case and the inset (b) demonstrates it for the nearest neighbour coupling case clearly highlighting the consequence of long-range connectivity. {Even for small values of $g$ the first case (all-to-all) shows partial  localization whereas we see complete delocalization in the second case (nearest neighbour). }

Since the dimension $\mathcal{N}_d$ of the Hilbert space is very large it is evident that simulating this system with exact quantum numerics for larger $N$ is essentially impossible. Therefore, to analyse the behaviour of large networks, we resort to semi-classical approximation where we decouple correlation functions such as, $\langle a_i\sigma_i^z  \rangle \approx  \langle a_i\rangle \langle \sigma_i^z\rangle$. When feasible, we have benchmarked semi-classical results with direct quantum simulations thereby supporting the usage of semi-classical approach \cite{SM}. Introducing the definitions $\langle a_i \rangle \equiv \alpha_i$, $\langle \sigma_i^-\rangle \equiv \beta_i$ and $\langle \sigma^z_i \rangle \equiv w_i$, the semi-classical equations of motion are given by

\begin{eqnarray}
      \dot{\alpha}_i &=& -i \omega_c \alpha_i +iJ \sum_{\substack{j  = 0 \\(j \neq i)}}^{N-1}\alpha_j -ig \beta_i, \\
      \dot{\beta}_i &=& -i\omega_q\beta_i + ig \alpha_i w_i, \,\,
      \dot{w}_i = 2ig \left( \alpha_i^*\beta_i -\alpha_i \beta_i^*\right).
\end{eqnarray}
Such an approximation typically fails for small photon numbers where quantum fluctuations play a significant role. In Fig.~\ref{fig:heatmap} (a), we show the heat map for the degree of localization as one varies the light-matter interaction strength ($g$) and the range of connectivity of the network ($D$). We see that for a given range of connectivity there exists a critical value of $g$ which demarcates the delocalization-localization transition. The results presented here are for the resonant case ($\omega_c = \omega_q$) and therefore correspond to the case of {strongly} anharmonic cavity networks. Our results are generalizable to the off-resonant JC case ($\omega_c \neq \omega_q$).  A well-known limit is the dispersive JC case where the detuning between the cavity and qubit frequencies is large compared to light-matter interactions ($g$). In this limit, the system can map to an attractive or repulsive BH  network \cite{boissonneault2009dispersive}. Irrespective of this connection between dispersive JC and BH network, the BH network with varied connectivity is a fascinating many body system in itself and warrants a thorough investigation which is done next. 

\textit{Network with Bose-Hubbard (BH) nonlinearity:}\label{section 4}
The governing Hamiltonian for the system is same as Eq.~\eqref{Hamil} with the $i^{\rm th}$ unit described by the Hamiltonian $H_i=\omega_c a^{\dagger}_ia_i-\frac{U}{2}n^2_i$. Here $U$ quantifies the strength of on-site attractive interaction. Our analysis involves networks with large number of cavity units and large photon numbers. Tackling such large-scale system is beyond the scope of fully quantum treatment due to numerical complexity. Considering the success of semiclassical theory in analyzing BH systems 
\cite{zibold2010classical,milburn1997quantum,dey2017self,chuchem2010quantum,smerzi1997quantum,albiez2005direct,SM} we employ semi-classical approximation to our system and write the EOM as
\begin{eqnarray}
\dot{\alpha}_i &=& -i\omega_c {\alpha}_i+iJ\sum_{\substack{j  = 0 \\(j \neq i)}}^{N-1}{\alpha}_j+i U |\alpha_i|^2 \alpha_i,
\label{eom_bh}
\end{eqnarray}
where $\langle a_i \rangle \equiv \alpha_i$. We initiate the system by populating just the test unit, i.e.,  $\alpha_0=\sqrt{10}$ with the rest of them taken to be empty, i.e., $\alpha_{j\neq 0}=0$. 

\begin{figure}[h]
  \centering
  \includegraphics[scale = 0.38]{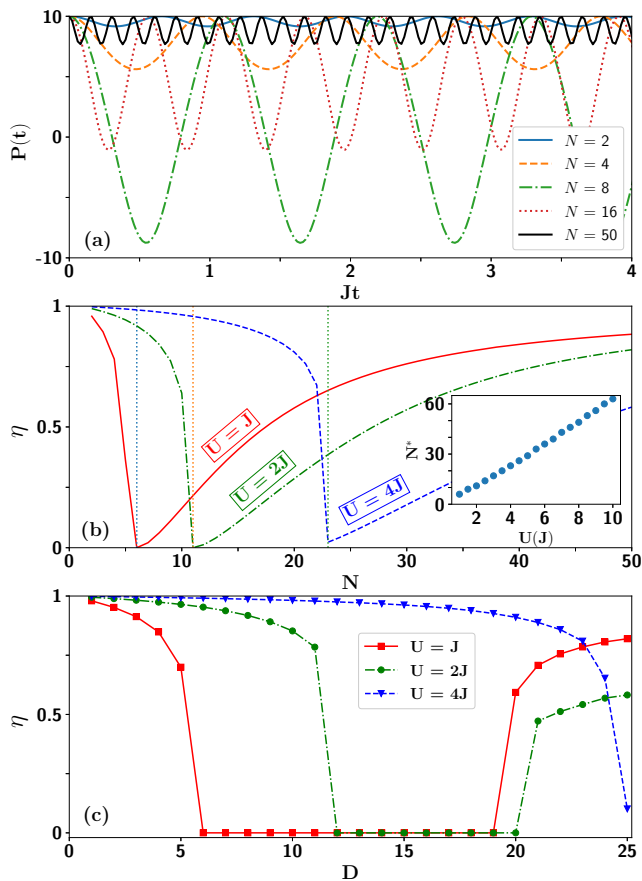}
  \caption{(a)  $P(t)$ as a function of time (in units of $1/J$) when uniform onsite nonlinearity (all-to-all BH network) is present in each cavity with $U = J$ for various values of $N$. 
  (b) Degree of localization versus $N$ for various values of BH nonlinearity. The dashed vertical lines mark the threshold number $N^{*}$ after which the network resembles a linear behaviour. Inset in (b) shows the behaviour of $N^{*}$ as a function of BH nonlinearity ($U$). (c) Degree of localization as a function of number of neighbours ($D$) for various values of $U$ ($N=50, N_{\rm p} = 20$).}
  \label{fig1_bh}
\end{figure}

With these initial conditions, we numerically solve the set of coupled nonlinear differential equations [Eq.~\eqref{eom_bh}] and plot $P(t)$ in Fig.~\ref{fig1_bh} (a).
We observe that the localization deteriorates as the number of units $N$ increases from $2$ to $8$. This feature is exactly opposite to the linear cavity network discussed earlier. Interestingly, with further increase of $N$ from $8$ to $50$, Fig.~\ref{fig1_bh} (a) shows enhancement of localization. In other words, beyond certain $N$ (say $N^{*}$) the system behaves similar to an almost linear network (the effect of $U$ is of course still there). This non-monotonic feature is clearly {visible} in Fig.~\ref{fig1_bh} (b) for three different values of $U$. The vertical dashed lines mark the $N^{*}$ where the system transits from nonlinear to almost linear in terms of localization. It is also important to note that the transition point $N^{*}$ moves towards higher values as $U$ is increased and the detailed relationship between $N^{*}$ and $U$ is captured in the inset of Fig.~\ref{fig1_bh} (b). We further investigate the role of connectivity in Fig.~\ref{fig1_bh} (c) where we plot the degree of localization $\eta$ as a function of D for $N = 50$ units and $N_{\rm p} = 20$ photons. For lower values of $U$ ($ = J, \ 2J$), we observe that $\eta$ initially decreases to $0$ and then again increases to its all-to-all connectivity value after being completely delocalized over a range of $D$. For higher values of $U$ ($= 4J$), the system is never in the completely delocalized state, rather it falls from complete localization to its all-to-all connectivity value. The initial fall in $\eta$ takes place due to the increase in the number of pathways for the photons to escape, but for higher values of $D$, the system closely resembles the all-to-all configuration and thus $\eta$ increases again. For higher $U$, the effect of connectivity is dominated over by the onsite attractive interaction. In {Fig.~\ref{fig:heatmap} (b)} we investigate the role of the number of neighbours $D$ and $U$ on $\eta$ for $N = 50$ and $N_{\rm p} = 50$. A clear transition from delocalized to self-trapped regime is observed and there is a linear relationship between the value of $U$ at which the transition takes place and the number of neighbours.

The nonlinear cavity in all-to-all network contains rich physics of competing factors such as $U$ and $N$. On the one hand, increasing the number of units $N$ increases the number of pathways for the movement of photons. On the other hand, the BH interaction $U$ restricts photon flow due to onsite photon-photon attraction. This is the reason that degree of localization decreases when one increases $N$ for a fixed $U$. However, after crossing some threshold $N^{*}$ the degree of localization {improves} because the system closely resembles a linear system. Needless to mention, it is paramount to establish that this non-monotonic behaviour is not an artefact of semi-classical approximation [Eq.~(\ref{eom_bh})]. To do so, when feasible we perform an exact quantum calculation and demonstrate the non-monotonic feature \cite{SM}. 

\textit{Conclusion:}
 Macroscopic self-trapping (or lack thereof) is a phenomenon that has been a subject of intense investigation in various physical platforms. We fill a major gap in this direction by studying  localization physics in scalable circuit-QED networks with connectivity that can be varied from finite-range to fully long-range.  {For the case of harmonic networks we demonstrate localization that is rooted in all-to-all connectivity as opposed to disorder \citep{anderson1958absence} or interactions \citep{mott1968metal, albiez2005direct}}. After providing analytical results for the harmonic network, we unveiled the exotic interplay among anharmonicity/interactions, kinetic hopping and network connectivity that often leads to counter-intuitive behaviour. We present exact quantum numerics for feasible system sizes and study large scale networks using semi-classical approach. When possible, we do a comparative study between quantum and semi-classical approaches \cite{SM}. {Bearing in mind realistic experimental setups, we investigate the role of imperfections in cavity/qubit and disorder. Our open quantum system computations  \cite{SM} show that for experimentally accessible time scales one can still capture the interesting results of localization. We also demonstrate robustness to disorder  \cite{SM}.}

Our findings are relevant for physical systems with long-range connectivity. In recent experiments related to large quantum computational architectures, designing nontrivial geometry as well as engineering connectivity through multiple qubits have become of increasing focus \cite{song2019generation, xu2020probing, cross2019validating, onodera2020quantum, hazra2021ring}. Our findings are potentially experimentally realizable in existing circuit-QED platforms and are expected to play a pivotal role for exploring other setups with non-trivial geometries.  As a future direction, it would be interesting to consider driven-dissipative quantum networks (with varied connectivity) and investigate their non-equilibrium steady-state properties. It is challenging and interesting to explore level spacing statistics (and spectral transitions) in these networks and this is expected to have a deep connection to the localization/delocalization phenomena \cite{prasad2021dissipative}. Our work can be further extended to more interesting geometries like hyperbolic lattices \cite{kollar2019hyperbolic,kollar2020line}. Such exotic deformations of lattices have been realised in experiments using coplanar waveguide  resonators \cite{kollar2019hyperbolic}, and it would be interesting to investigate the effect of curvature along with connectivity on such networks.\\

\textit{Acknowledgements:}
We thank R. Vijayaraghavan, S. Hazra, D. O'Dell and H. K. Yadalam for useful discussions. MK acknowledges the support of the Ramanujan Fellowship (SB/S2/RJN-114/2016), SERB Early Career Research Award (ECR/2018/002085) and SERB Matrics Grant (MTR/2019/001101) from the Science and Engineering Research Board (SERB), Department of Science and Technology, Government of India. MK acknowledges the support of the Department of Atomic Energy, Government of India, under Project No. RTI4001.  We gratefully acknowledge the ICTS-TIFR high performance computing facility. MK thanks the hospitality of  \'Ecole Normale Sup\'erieure (Paris).

\bibliography{bibliography.bib}

\end{document}


	
	
\date{\today}
\newcommand{\titlename}{Supplementary material for `Localization and delocalization in networks with varied connectivity'}

\title{\titlename}
	
\author{Tamoghna Ray}
\email{tamoghna.ray@icts.res.in} 
\address{International Centre for Theoretical Sciences, Tata Institute of Fundamental Research, Bengaluru -- 560089, India}

\author{Amit Dey}
\email{amit.dey.85@gmail.com }
\address{Ramananda College, Bankura University, Bankura 722122, India}
\address{International Centre for Theoretical Sciences, Tata Institute of Fundamental Research, Bengaluru -- 560089, India}

\author{Manas Kulkarni}
\email{manas.kulkarni@icts.res.in}
\address{International Centre for Theoretical Sciences, Tata Institute of Fundamental Research, Bengaluru -- 560089, India}

\maketitle
	
\tableofcontents 

\vspace{0.8cm}

\section{Analytical Results for Harmonic networks}
	
In this section, we will derive analytical expressions for $P(t)$ in both the finite-range [Sec.~\ref{Sec_FR}] and all-to-all coupling [Sec.~\ref{Sec_a2a}] case when the units $H_i$ are harmonic oscillators, i.e., $H_i = \omega_c a^\dagger_i a_i$ . We shift to the Fourier space, 
\begin{equation}
a^\dagger_j = \frac{1}{\sqrt{N}}\sum_{k=0}^{N-1} e^{-i \frac{2\pi}{N} k j}a^\dagger_k,
\label{FT}
\end{equation}
where $j =0,1,2 \dots N-1$ is the lattice index and $k$ represents the wave number. Substituting Eq.~\eqref{FT} in the unit Hamiltonian we get,
\begin{equation}
\sum_{j=0}^{N-1}H_{j} = \frac{\omega_c}{N}\sum_{\substack{j  = 0 \\k,k'=0}}^{\substack{j  = N-1 \\k,k'=N-1}} a^\dagger_k a_{k'} e^{-i\frac{2\pi}{N}(k - k')j}.
\end{equation}
Now using the Fourier decomposition of the Kronecker delta,
\begin{equation}
\frac{1}{N}\sum_{j=0}^{N-1} e^{-i\frac{2\pi}{N}(k - k') j}  = \delta_{kk'},
\label{delta_iden}
\end{equation}
we get rid of the sum over $j$ and contract the sum over $k'$ which finally yields, 
\begin{equation}
\therefore \sum_{j=0}^{N-1}H_{j} = \omega_c\sum_{j=0}^{N-1} a^\dagger_j a_j = \omega_c\sum_{k=0}^{N-1} a^\dagger_k a_k.
\label{SHO_FT}
\end{equation}

Now, to evaluate the coupling Hamiltonian, we will require to simplify terms such as $\sum_{j=0}^{N-1}(a^{\dagger}_ja_{j+d} + h.c)$ where $d$ is some integer. Performing a Fourier transform gives,	
\begin{equation}
\begin{gathered}
\sum_{i=0}^{N-1}(a^\dagger_i a_{i+d} + h.c.) = \frac{1}{N} \sum_{\substack{i  = 0 \\k,k'=0}}^{\substack{i  = N-1 \\k,k'=N-1}} \big(e^{-i \frac{2\pi}{N}k i + i \frac{2\pi}{N}k' (i+d)}\times \\a^\dagger_k a_{k'} + h.c. \big).			
\end{gathered}
\end{equation}

Again, using Eq.~\eqref{delta_iden} and contracting the sum over $k'$ we get,
\begin{equation}
\sum_{i=0}^{N-1}(a^{\dagger}_i a_{i+d} + h.c) = 2\sum_{k=0}^{N-1}a^\dagger_k a_k \cos \left(\frac{2\pi}{N}k d\right).
\label{coup_FT}
\end{equation}

\begin{figure}[h]
\centering
\includegraphics[scale = 0.4]{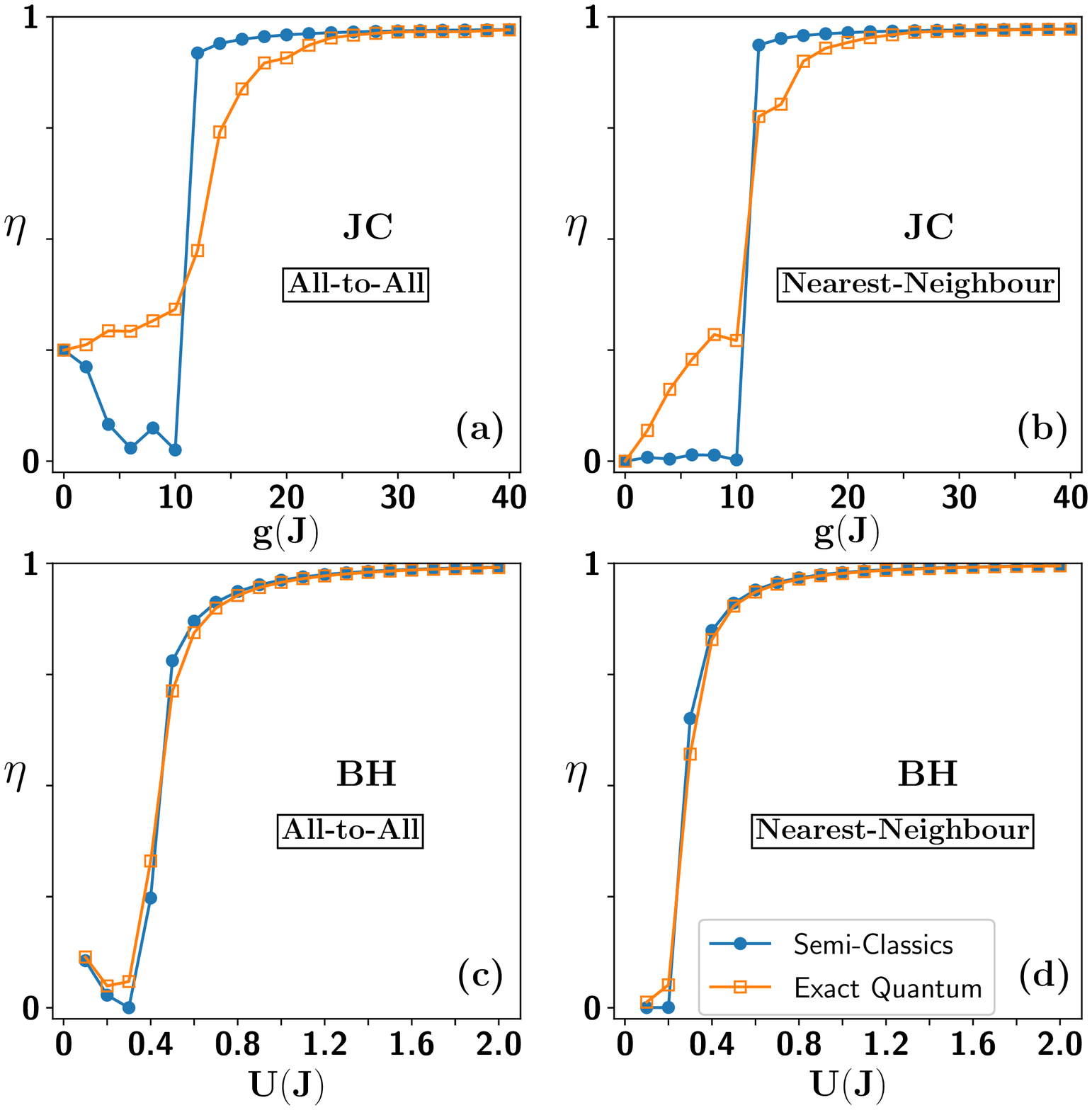}
\caption{Comparison of exact quantum numerics and semiclassical approach. (a) and (b) show $\eta$ as a function of $g$ for the JC network with $N = 4$ and $N_{\rm p} = 20$, for all-to-all and nearest neighbour coupling respectively. (c) and (d) show $\eta$ as a function of $U$ for the BH network with $N = 4$ and $N_{\rm p} = 20$, for all-to-all and nearest neighbour coupling respectively.}   
\label{quantum_vs_semi}
\end{figure}


\subsection{Finite-Range Coupling}
\label{Sec_FR}

For finite-range coupling ($D <\lceil N/2\rceil$), using Eq.~\eqref{coup_FT}, the coupling part of the Hamiltonian ($H_{\rm coupling}$)  becomes,

\begin{eqnarray}
H^{\rm finite-range}_{\rm coupling} &=& -J\sum_{i=0}^{N-1}\sum_{d=1}^{D}(a^{\dagger}_ia_{i+d} + h.c) \nonumber \\ &=& -2J\sum_{k=0}^{N-1}\sum_{d = 1}^{D} \cos\left(\frac{2\pi}{N}kd \right) a_k^\dagger a_k \nonumber \\
&=& -2JD a^\dagger_0a_0 - 2J\sum_{k=1}^{N-1}f(k)a^{\dagger}_ka_k,
\end{eqnarray}
where
\begin{equation}
f(k) = \sum_{d = 1}^{D} \cos\left(\frac{2\pi}{N}kd \right)  = \frac{\cos\left( \frac{D+1}{N}\pi k\right)\sin\left(\frac{D\pi k}{N} \right)}{\sin \left( \frac{\pi k}{N} \right)}.
\end{equation}
Using $\dot{a_k} = i[H,a_k]$, where $H$ is in the momentum basis, the equations of motion become,	
\begin{equation}
\dot{a}_k =\begin{cases}
-i\omega_c a_k +i2J f(k)a_k & \text{$k \neq 0$}\\
-i\omega_c a_k +i2JDa_k & k=0.
\end{cases}
\label{eq:akdot}
\end{equation}
Note that in above Eq.~(\ref{eq:akdot}), we treat the $k \to 0$ limit carefully and separately. Solution to Eq.~(\ref{eq:akdot}) reads,	
\begin{equation}
a_k(t) = \begin{cases}
a_k(0)\exp\Big[-i\Big(\omega_c -2Jf(k)\Big)t\Big] & \text{$k \neq 0$}\\
a_k(0)\exp[-i(\omega_c -2DJ)t] & k=0.
\end{cases}
\label{sol_finite_dft}
\end{equation}
The initial values $a_k(0)$ are determined by $a_j(0)$ using Eq.~\eqref{FT},	
\begin{equation}
a_k(0) = \frac{1}{\sqrt{N}} \sum_{j=0}^{N-1} e^{-i k j}a_j(0). 
\end{equation}
Now we initiate the system in a way such that there are $N_{\rm p}$ bosons in the test unit and all other sites are empty, i.e., $\langle a_0(0)\rangle = \sqrt{N_{\rm p}}$ and $\langle a_{j \neq 0}(0)\rangle = 0$. Therefore, we have, 
\begin{equation}
\langle a_k(0) \rangle = \sqrt{\frac{N_{\rm p}}{N}}.
\label{ak_init}
\end{equation}
The population difference operator $P(t)$ can be written as
\begin{equation}
P(t) = 2n_0(t) - \sum_{j = 0}^{N-1} n_j(t),
\end{equation}
where we recall that $n_i (t) = \langle a_i^\dagger (t) a_i  (t) \rangle$. Since there are no dissipative processes involved, we have $\sum_{j = 0}^{N-1}n_j(t) = N_{\rm p}$. Using Eq.~\eqref{FT} we get,
\begin{eqnarray}
\label{dft1}
P(t) &=& \frac{2}{N}\sum_{k,k'=0}^{N-1}a^\dagger_ka_{k'} - N_{\rm p} \nonumber \\
&=& \frac{2}{N}\Bigg\{a^\dagger_0(t)a_0(t) + \sum_{k=1}^{N-1}a^\dagger_k(t)a_0(t)  \\ &+& \sum_{k=1}^{N-1}a^\dagger_0(t)a_k(t)  + \sum_{k,k'=1}^{N-1}a^\dagger_k(t)a_k'(t)   \Bigg\} - N_{\rm p} \nonumber
\end{eqnarray}

Substituting Eq.~\eqref{sol_finite_dft} we get,

\begin{eqnarray}
P(t) &=& \frac{2N_{\rm p}}{N^2}\Bigg\{ 1 + \sum_{k,k'=1}^{N-1}\exp\left[i2Jt\left( f(k') - f(k) \right)\right] \nonumber \\ &+& 2\sum_{k=1}^{N-1}\cos\left[  2Jt\left(f(k) -D \right) \right] \Bigg\}- N_{\rm p}.
\label{finite_P}
\end{eqnarray}


\subsection{All-to-All Coupling}
\label{Sec_a2a}
For all-to-all coupling and for $D \geq \lceil N/2\rceil$ we follow the same steps as in the case of finite-range, with the necessary changes in the coupling term.

\begin{eqnarray}
H^{\rm all-to-all}_{\rm coupling}&=&	-\frac{J}{2}\sum_{\substack{i,j  = 0  \\ i \neq j}}^{N-1} \left(a^\dagger_i a_j + h.c.\right) \nonumber \\ &=& -J\sum_{k=0}^{N-1}\sum_{d = 1}^{N-1} \cos\left(\frac{2\pi}{N}kd \right) a_k^\dagger a_k \nonumber \\
&=& -J(N-1)a^{\dagger}_0a_0 \nonumber \\ &-& J\sum_{k=1}^{N-1} \frac{\cos\left(\pi k\right)\sin\left(\frac{N-1}{N}\pi k\right)}{\sin\left(\frac{\pi k}{N}\right)}a^\dagger_k a_k. \nonumber \\ 
\end{eqnarray}

The equations of motion are given by,
\begin{equation}
\dot{a}_k =\begin{cases}
-i\omega_c a_k -iJa_k, & \text{$k \neq 0$}\\
-i\omega_c a_k +iJ(N-1)a_k & k=0,
\end{cases}
\end{equation}
whose solution reads, 	
\begin{eqnarray}
a_k(t) = \begin{cases}
a_k(0)e^{-i(\omega_c +J)t} & \text{$k \neq 0$}\\
a_k(0)e^{-i(\omega_c -(N-1)J)t} & k=0.
\end{cases}
\label{sol_a_to_a_dft}
\end{eqnarray}
Substituting Eq.~\eqref{sol_a_to_a_dft} in Eq.~(\ref{dft1}) we get, 
\begin{eqnarray}
\label{eq:pta2a_sm}
P(t) &= &\frac{2}{N}\Bigg\{ \frac{N_{\rm p}}{N} + \frac{N_{\rm p}}{N}(N-1)^2 \nonumber \\ &+& \frac{N_{\rm p}}{N}(N-1) 2\cos(NJt) \Bigg\} - N_{\rm p}.
\end{eqnarray}


Rearranging the terms in Eq.~(\ref{eq:pta2a_sm}) we get,
\begin{equation}
P(t) = \frac{N_{\rm p}}{N^2}[1 + (N-1)(N + 4\cos (NJt) -3)].
\end{equation}


\begin{figure}[h]
\centering
\includegraphics[width=3.in]{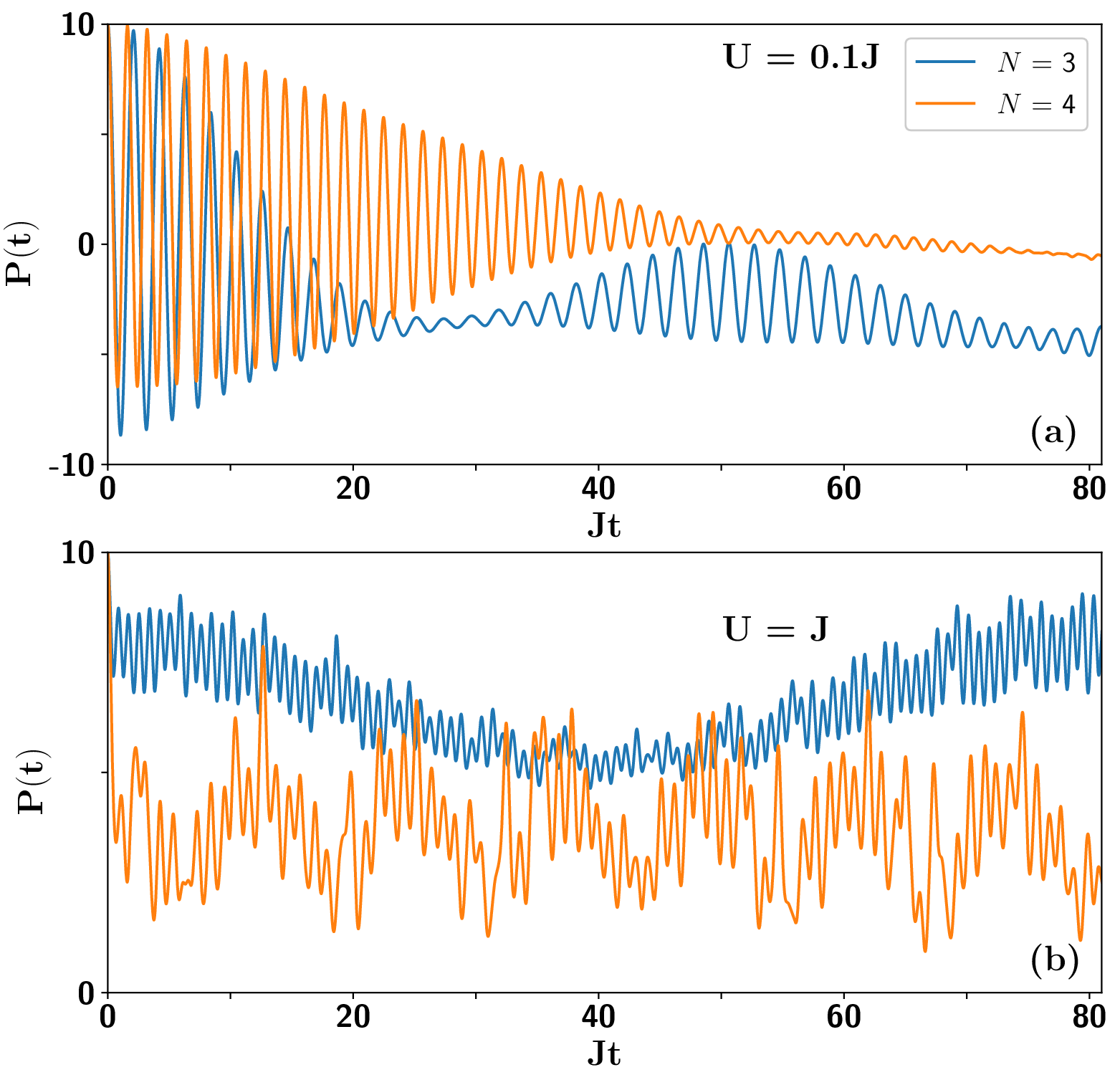}
\caption{Results of fully quantum treatment for BH network ($N=3,4$ and $N_{\rm p} = 10$) with all-to-all connectivity. Population difference for (a) $U=0.1 J$ and (b) $U=J$ as a function of time (in units of $1/J$). It is clear that for $U=0.1J$ the case of $N=4$ is more localized whereas for $U=J$ the case of $N=3$ is more localized thereby showing non-monotonicity in the quantum case. }  
\label{fig2_bh}
\end{figure}

\begin{figure}[h]
\centering
\includegraphics[scale = 0.4]{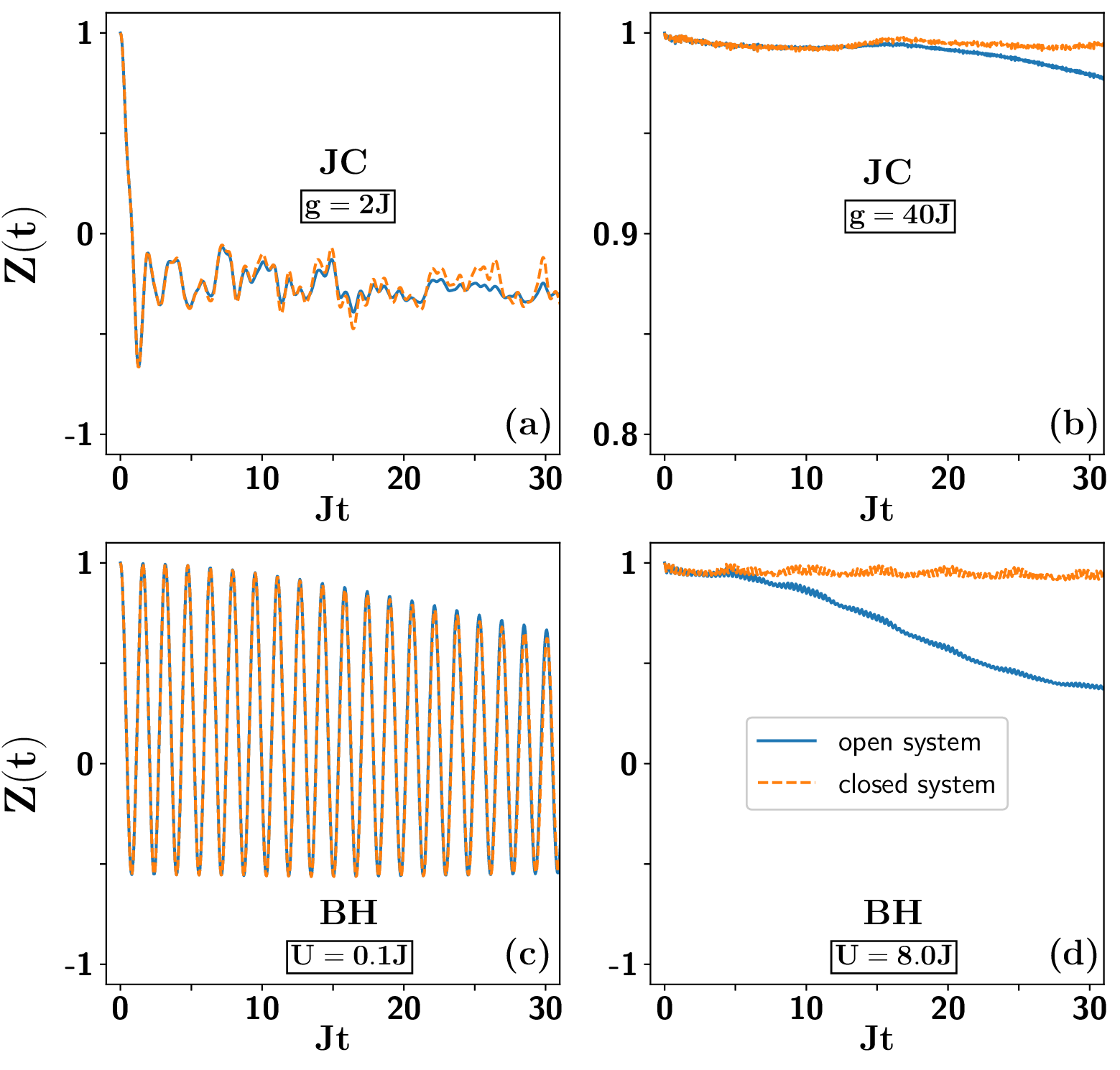}
\caption{Comparison of exact quantum numerics for closed and open systems. (a) and (b) show the evolution of $Z(t)$ defined in Eq.~(\ref{eq:zt}) as a function of time (in units of $1/J$) for the JC network with $N = 3$, $N_{\rm p} = 5$, $\kappa = 0.01J$, $\gamma = 0.01J$, for $g = 2J$ and $g = 40J$ respectively. (c) and (d) show the evolution of $Z(t)$ as a function of time (in units of $1/J$) for the BH network for $U = 0.1J$ and $U = 8J$ respectively ($N = 4$, $N_{\rm p} = 4$, $\kappa = 0.01J$).}   
\label{open_vs_closed}
\end{figure}

\section{Comparison between exact Quantum computations and semi-classical approach}

This section is dedicated to comparing (whenever feasible) exact quantum computations with the semi-classical approach. This is of paramount importance since both the JC and the BH networks are nonlinear systems and hence doing exact quantum calculations become unfeasible beyond a point. Therefore, it is important to have a benchmark so that one can use the semi-classical approach for larger scale problems. As mentioned in the main text, the dimension $\mathcal{N}_d$ of the Hilbert space for the entire system of JC network is extremely large and is given by 
$\mathcal{N}_d = \left[2 (N_{\rm p}+1)\right]^{N}$.
For e.g., when the number of units is  $N=5$ and if we take $N_{\rm p} = 10$ photons then $\mathcal{N}_d = 22^5$ (which is larger than $2^{22}$). This means that beyond a few number of units, the exact quantum calculations become unfeasible and one needs to resort to other approaches such as the semi-classical method. 

In Fig.~\ref{quantum_vs_semi}, we demonstrate the agreement between exact quantum numerics and semi-classical results for the JC and BH networks both for finite-range and all-to-all coupling. In particular, we investigate the degree of localization as a function of the strength of the nonlinearity. We find that even the quantitative agreement is reasonably good.

As a consequence of this agreement between semi-classical and exact quantum methods, we elaborate on the non-monotonic behaviour of localization for the BH network that was shown in Fig.~5 (b) of the main text. A natural question is whether this non-monotonic behaviour ($\eta$ versus $N$) is a consequence of semi-classical approximation or whether it is also present in fully quantum calculations. 
In Fig.~5 (b) of the main text, we have shown the non-monotonic behavior of localization for BH nonlinearity by considering semi-classical theory. Here, we will show that this non-monotonic behavior can also be verified for the quantum case. 
We observe in Fig.~5 (b) of the main text that,  in between the first two vertical dashed lines, localization for the $U=J$ case is better when $N$ is larger (the solid red line was curving upwards). On the other hand, for $U = 2J$, the degree of localization decreases when $N$ increases (the green dashed-dotted line was curving downwards). This is rooted in the fact that, upon increasing $N$, the $U=J$ case enters the linear-like region earlier than the $U = 2J$ case. This implies that for two values of $N$, say $N=3$ and $N=4$, the saturation values of $P(t)$ would be in opposite trend for $U=J$ and $U=2J$. We find this also to be true when we perform the exact quantum calculations. 
Comparing Figs. \ref{fig2_bh} (a) and (b) we see that the population difference $P(t)$ for $N=4$
case lies above the $N = 3$ case when $U = 0.1J$, whereas the order is reversed for $U =J$ case. Therefore, this feature is consistent with the observation we have for the semi-classical case and is a reflection of non-monotonicity.


\begin{table}[h]
\centering
\begin{tabular}{ p{2cm}p{4cm}  }
\hline
\multicolumn{2}{c}{Typical range of parameter values ($\times 2\pi$) } \\
\hline \hline
$\omega_c$   &   4 - 8  GHz \\
$\omega_q$   & 4 - 8  GHz\\
$g$   & 1 - 300 MHz\\
$U$ & 150 - 350 MHz\\
$J$& 50 kHz - 50 MHz\\
$\kappa$&50 kHz - 50 MHz\\
$\gamma$&20 - 200kHz\\
$\gamma_\phi$& 20 - 200kHz\\

\hline

\end{tabular}  
\caption{Summary of typical range of parameter values obtainable in circuit-QED \cite{koch2007charge, raftery2014observation, houck2012chip, song2019generation,naik2017random}. }    
\label{tab:parameters}
\end{table}


\begin{figure}
\centering
\includegraphics[scale = 0.4]{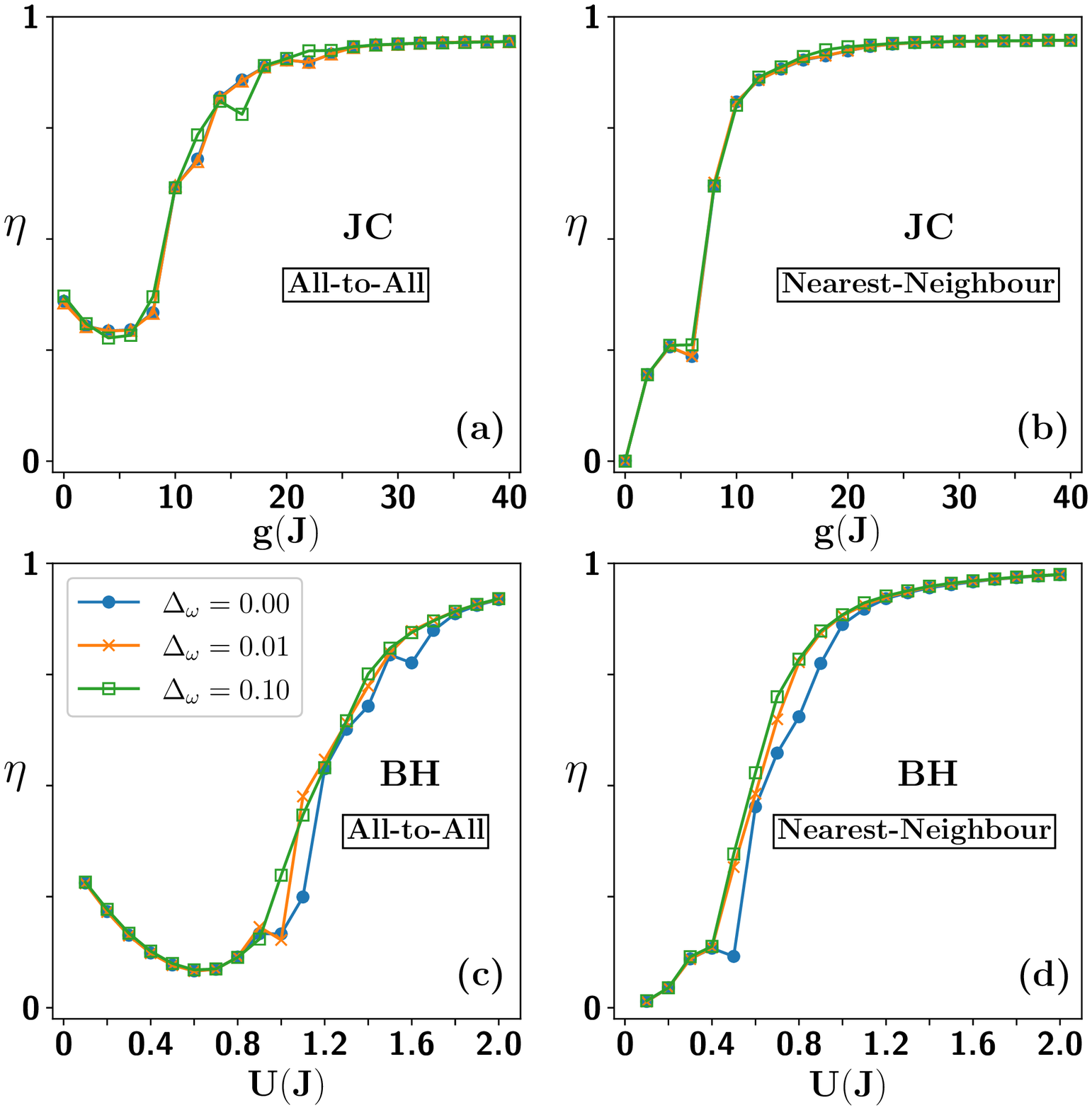}
\caption{Effect of disorder characterized by $\Delta_\omega$ on degree of localization for exact quantum numerical solution. Results for $\Delta_\omega = 0.00$, $0.01$ and $0.10  \omega_c$ are compared. (a) and (b) show $\eta$ as a function of $g$ for the JC network with $N = 5$ and $N_{\rm p} = 10$, for all-to-all and nearest neighbour coupling respectively. (c) and (d) show $\eta$ as a function of $U$ for the BH network with $N = 5$ and $N_{\rm p} = 10$, for all-to-all and nearest neighbour coupling respectively.}   
\label{disorder}
\end{figure}

\begin{figure}
\centering
\includegraphics[scale = 0.4]{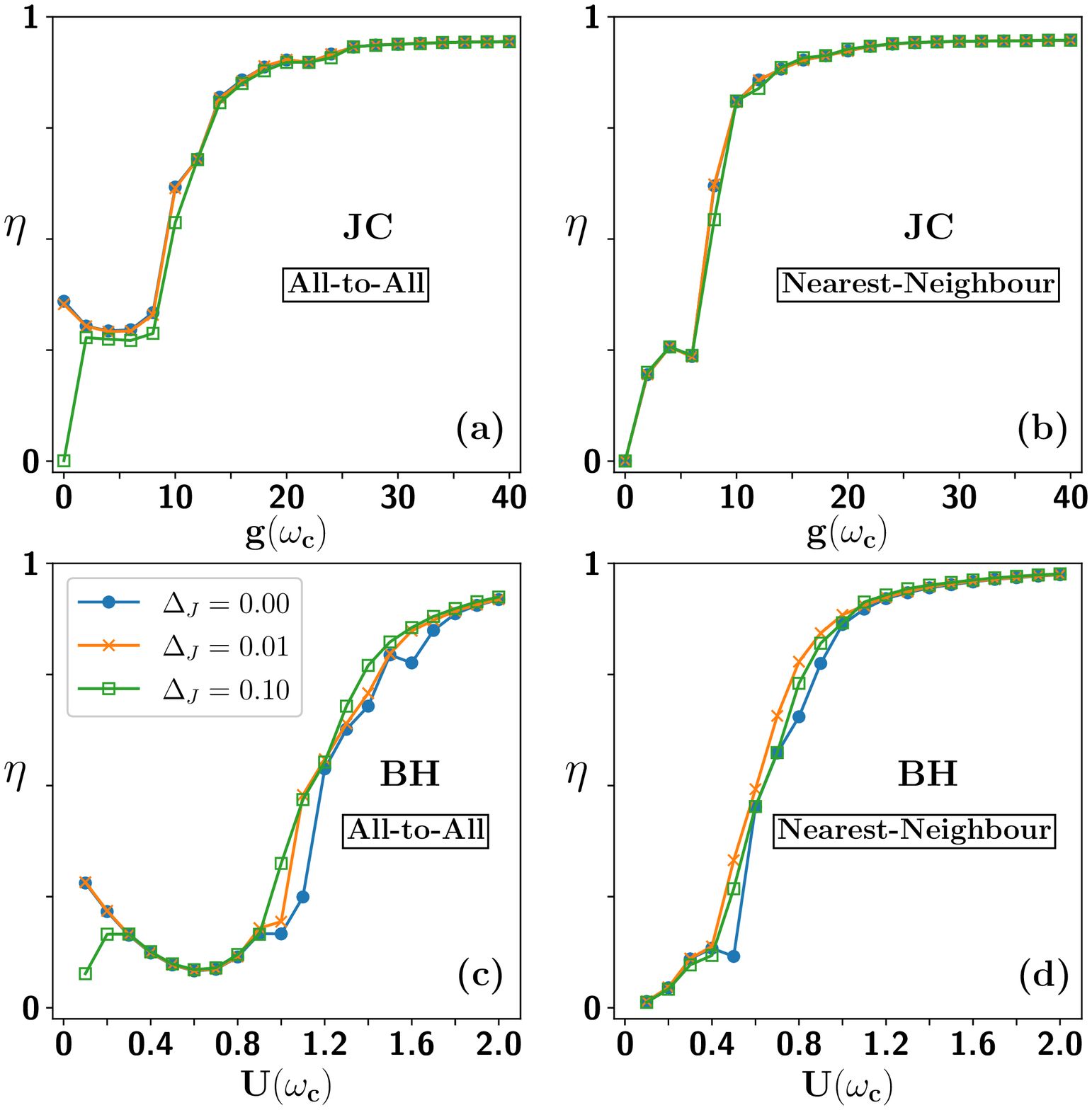}
\caption{Effect of disorder of the kinetic hopping $J$ characterized by $\Delta_J$ on degree of localization for exact quantum numerical solution. Results for $\Delta_J = 0.00, 0.01$ and $0.10  J$ are compared. (a) and (b) show $\eta$ as a function of $g$ (in units of $\omega_c$) for the JC network with $N = 5$ and $N_{\rm p} = 10$, for all-to-all and nearest neighbour coupling respectively. (c) and (d) show $\eta$ as a function of $U$ (in units of $\omega_c$) for the BH network with $N = 5$ and $N_{\rm p} = 10$, for all-to-all and nearest neighbour coupling respectively.}   
\label{Jdisorder}
\end{figure}

\section{Closed versus Open Quantum Systems}
In this section, we investigate the consequence of imperfections due to coupling with the environment which is inevitable in physical systems. Examples of such imperfections include dissipation in the cavity and qubit decay/dephasing. 
In the main text, we neglected environmental effects to focus mainly on the interplay between nonlinearity, kinetic hopping and connectivity. Here, we discuss the case when we have open quantum systems. The range of parameters that we consider are motivated by experiments and summarized in Table.~\ref{tab:parameters}. 
To incorporate environmental effects in the system dynamics we rely on the Linblad formalism and exploit  the following local Linblad quantum master equation. 
\begin{eqnarray}
\begin{gathered}
\dot{\rho}(t)=-i[H,\rho(t)]+\kappa\sum_{j=0}^{N-1} \mathcal{L}[a_j]+\gamma \sum_{j=0}^{N-1} \mathcal{L}{[\sigma^-_j}]\\ + \gamma_\phi\sum_{j=0}^{N-1} \mathcal{L}{[\sigma^z_j}],	
\end{gathered}
\label{linblad}
\end{eqnarray}
where $\mathcal{L}[A_j]=\big(2A_j\rho(t)A^{\dagger}_j-A^{\dagger}_j A_j\rho(t)-\rho(t)A^{\dagger}_jA_j) \big/2$. $\rho(t)$ is the reduced density matrix of the system, which is obtained by tracing out the environmental degrees of freedom from the full density matrix (of the system and the environment). It is important to note that we restrict ourselves to local Lindblad description although, in general, a rigorous derivation starting from a microscopic system-bath Hamiltonian is expected to lead to various variants of quantum master equations~\cite{purkayastha2016out} depending on the parameter regimes. Furthermore, we consider very low temperature ($T$) regime and neglect the effect of thermal excitations in Eq.~\ref{linblad}. This is because typical temperature regimes in experiments range from $T \sim 10 - 30 \text{ mK}$. For BH networks, qubit degrees of freedom are absent and the governing Lindblad master equation can be obtained by putting $\gamma=\gamma_\phi= 0$ in Eq.~(\ref{linblad}). Since, in the open system case, the total number of excitations is not conserved, we introduce a slightly modified variant of $P(t)$ which is 
\begin{equation}
Z(t) = \frac{P(t)}{ \sum_{j = 0}^{N-1} n_j(t)}. 
\label{eq:zt}
\end{equation}
$Z(t)$ gives the ratio of the imbalance to the total number of excitations at any given time. In Fig.~\ref{open_vs_closed}, this is plotted both for the JC (top panel) and BH network (bottom panel). We notice that self-trapping in both cases (right panels), persists for reasonably long times after which there is a decay of $Z(t)$. 

%

\section{Robustness to disorder}
The role of disorder due to error margins in parameter values of the network is discussed in this section. In experiments, it is almost impossible to design network elements such as cavities and qubits of a given frequency or level gap without any error margin. For example, in a typical experiment, to design a cavity of frequency $\omega_c = 7 \times 2\pi \text{ GHz}$, there is an error margin of $\pm 70 \times 2\pi \text{ MHz}$ ($0.01 \omega_c$). In the main text, we have neglected such disorders in the network and one might wonder the consequence of disorder on the phenomenon of self-trapping. Thus, for our theoretical findings to be observed in potential experimental setups, it is important to check their robustness upon the inclusion of disorder. In order to proceed, we modify the cavity Hamiltonian in the following manner, 
\begin{equation}
H_{\rm {cavity}}^{\rm {disordered}} = \sum_{i = 0}^{N-1} ( \omega_{c_i} + \epsilon_i )  a_i^\dagger a_i 
\end{equation}
where 
$\epsilon_i$ is drawn from a uniform distribution ranging from $-\Delta_\omega$ to $\Delta_\omega$. In Fig.~\ref{disorder}, we plot the degree of localization $\eta$ as a function of the light-matter interaction $g$ for the JC network and the onsite attractive potential $U$ for the BH network. We present exact quantum numerical results for all-to-all and nearest neighbour connectivity with $\Delta_\omega = 0.01 \text{ and } 0.1 \omega_c$. {The effect of disorder on the kinetic hopping strength $J$ is also investigated by considering the hopping between two units $i$ and $j$ to be $J_{ij} = J + \epsilon_{ij}$ where, $\epsilon_{ij} = \epsilon_{ji}$ is again drawn from a uniform distribution ranging from $-\Delta_J$ to $\Delta_J$. In Fig.~\ref{Jdisorder} we present the degree of localization for the JC and the BH networks with all-to-all and nearest neighbour coupling with $\Delta_J = 0.01 \text{ and } 0.1J$.} The robustness to disorder {in both cavity frequency and kinetic hopping} is clearly manifested in Fig.~\ref{disorder} and Fig.~\ref{Jdisorder}.

\bibliography{bibliography.bib}
